\documentclass[super,prb,
twocolumn,reprint,
groupedaddress,superscriptaddress,
amsfonts,amssymb,amsmath,floatfix,
showkeys,showpacs,a4paper]{revtex4-1}

\usepackage{CJK}

\usepackage{graphicx}
\usepackage[english]{babel}
\usepackage{hyperref}
\usepackage{upgreek}
\usepackage{natbib}
\usepackage{lineno}
\usepackage{color}

\begin{document}
\begin{CJK*}{GB}{} 

\title{Magnetic properties of rare earth and transition metal based perovskite type high entropy oxides}

\author{Ralf Witte}\email{ralf.witte@kit.edu}
\affiliation{Institute of Nanotechnology, Karlsruhe Institute of Technology, 
 76344 Eggenstein-Leopoldshafen, Germany}

\author{Abhishek Sarkar}
\affiliation{Institute of Nanotechnology, Karlsruhe Institute of Technology, 76344 Eggenstein-Leopoldshafen, Germany}
\affiliation{KIT-TUD-Joint Research Laboratory Nanomaterials, Technical University Darmstadt,  64287 Darmstadt, Germany}

\author{Leonardo Velasco}
\affiliation{Institute of Nanotechnology, Karlsruhe Institute of Technology, 76344 Eggenstein-Leopoldshafen, Germany}

\author{Robert Kruk}
\affiliation{Institute of Nanotechnology, Karlsruhe Institute of Technology, 
 76344 Eggenstein-Leopoldshafen, Germany}
 
\author{Richard A. Brand}
\affiliation{Institute of Nanotechnology, Karlsruhe Institute of Technology, 
 76344 Eggenstein-Leopoldshafen, Germany}
\affiliation{Faculty of Physics and Center for Nanointegration Duisburg-Essen (CENIDE), University of Duisburg-Essen, Lotharstr. 1, 47048 Duisburg, Germany}

\author{Benedikt Eggert}
\affiliation{Faculty of Physics and Center for Nanointegration Duisburg-Essen (CENIDE), University of Duisburg-Essen, Lotharstr. 1, 47048 Duisburg, Germany}

\author{Katharina Ollefs}
\affiliation{Faculty of Physics and Center for Nanointegration Duisburg-Essen (CENIDE), University of Duisburg-Essen, Lotharstr. 1, 47048 Duisburg, Germany}

\author{Eugen Weschke}
\affiliation{Helmholtz-Zentrum Berlin f\"ur Materialien und Energie (HZB), Albert-Einstein-Str. 15, 12489 Berlin, Germany}

\author{Heiko Wende}
\affiliation{Faculty of Physics and Center for Nanointegration Duisburg-Essen (CENIDE), University of Duisburg-Essen, Lotharstr. 1, 47048 Duisburg, Germany}

\author{Horst Hahn}
\affiliation{Institute of Nanotechnology, Karlsruhe Institute of Technology, 
 76344 Eggenstein-Leopoldshafen, Germany}
\affiliation{KIT-TUD-Joint Research Laboratory Nanomaterials, Technical University Darmstadt,  64287 Darmstadt, Germany}

\date{\today}
\begin{abstract}

High entropy oxides (HEO) are a recently introduced class of oxide materials, which are characterized by a large number of elements (i.e. five or more) sharing one lattice site  which crystallize in a single phase structure. One complex example of the rather young HEO family are the rare-earth transition metal perovskite high entropy oxides. In this comprehensive study, we provide an overview over the magnetic properties of three perovskite type high entropy oxides. The compounds have a  rare-earth site which is occupied by five different rare-earth elements, while the transition metal site is occupied by a single transition metal. In this way a comparison to the parent binary oxides, namely the orthocobaltites, -chromites and -ferrites  is possible. X-ray absorption near edge spectroscopy  (XANES),  magnetometry and M\"ossbauer spectroscopy are employed to characterize these complex materials. 


In general, we find surprising similarities to the magnetic properties of the binary oxides, despite the chemical disorder on the rare-earth site. However  distinct differences and interesting magnetic properties are also observed such as noncollinearity, spin reorientation transitions as well as large coercive fields of up to 2\,T at ambient temperature. Both the chemical disorder on the RE A-site, and the nature of the TM on the B-site play an important role in the physical properties of these high entropy oxides.

\end{abstract}
\pacs{ }
\maketitle
\end{CJK*}

\section{Introduction}
High entropy oxides (HEOs) are single-phase solid solutions which contain five or more cations in near-equiatomic amounts \cite{Rost2015,Berardan2016a,Sarkar2019}. Phase-purity in HEOs, despite the chemical complexity, along with homogeneous and statistical distribution of the cations on the respective Wyckoff sites make them rather appealing \cite{Rost2015,Rost2017a,Sarkar2019,Chellali2019}. Statistical distribution of the multiple equiatomic cations in HEOs essentially affects and increases the configurational entropy of the system (as calculated from the Boltzmann equation). Hence, the term "high entropy" is commonly used for describing multinary systems of these kinds following Murty et al. \cite{Murty2014}, wherein it has been proposed that any system with configurational entropy higher than 1.5 R (where R is the universal gas constant) can be grouped under this category. In addition, it is believed, and in some cases proven \cite{Rost2015, Berardan2016a, Sarkar2017} that the structural stability of HEOs is of often related to this increased configurational entropy. Further information pertaining to entropy based materials classification and related entropy-stabilization effects can be found elsewhere \cite{Rost2015,Miracle2017}.\cite{Sarkar2019}

The spectrum of HEOs has been expanded significantly with the discovery of several compositions with different crystal structures, such as rock-salt, fluorite, spinel and perovskite \cite{Rost2015,Gild2018,Dabrowa2018,Sharma2018b,Sarkar2017,Musico2019}. Given this compositional flexibility in HEOs, a broad range of composition-based property tailoring possibilities can be anticipated. Some examples of such properties in rock-salt type HEOs are for high room temperature Li$^{+}$ conductivity, highly reversible lithium storage capabilities, colossal dielectric constant, low thermal conductivities, tailorable magnetic transition temperatures, etc \cite{Berardan2016,Sarkar2018b,Berardan2016a,Braun2018,Meisenheimer2017,Zhang2019}. Unlike the other classes of HEOs, research activities on perovskite type HEOs (PE-HEOs) are only limited to few reports \cite{Jiang2018, Sharma2018b, Sarkar2017a, Witte2019a}.

Rare-earth transition metal based perovskites or their doped variants are one of the most renowned class of oxide perovskites. The interplay between the composition, the resulting structure and physical properties in these perovskites makes them suitable candidates for several engineering applications, exploiting properties  such as e.g., magneto-electric effects \cite{Molinari2017}, multiferroic effects\cite{Cheong2007}, catalytic activities\cite{Mawdsley2008}, electronic\cite{Goodenough2001}, electrochemical and related transport properties\cite{Koep2006,Skinner2003}. In case of PE-HEOs, either the A-site is substituted by multiple rare earth (RE) and/or the B-site is replaced by several different transitional metal (TM). Owing to this new design concept, it would be worth to explore the synergy between multiple cations in RE - TM perovskites and compare their properties and underlying principles with the parent ortho-cobaltites, -chromites and -ferrites.

In our previous study \cite{Witte2019a}, the magnetic properties of a group of RE - TM based PE-HEOs with multiple B-site TM cations were investigated. The presence of the multiple TM B-site cation in those group of PE-HEOs led to a complex magnetic characteristic, wherein competition between dominating antiferromagnetic (AFM) and small ferromagnetic (FM) interactions could be observed. In this study, the other group of PE-HEOs has been investigated where the B-site has been fixed to one TM cation but the A-site consists of several RE elements. A combination of element specific M\"ossbauer and X-ray absorption spectroscopy along with magnetometry has been used to understand the magnetic interaction in these complex systems.

\section{Experimental details}\label{sec:exp}

\subsection{Synthesis and structural characterization}\label{sec:exp_syn}

Aerosol based nebulized spray pyrolysis technique accompanied by subsequent heat treatment at 1200 $^{\circ}$C was typical procedure followed for the synthesis of PE-HEO powder. The detailed description of the synthesis route  is reported elsewhere \cite{Sarkar2017a}. 
Following up our previous report \cite{ Witte2019a}, in this study three systems (see below) with mixed RE A-site cations and a single TM B-site cation were investigated. 
\begin{itemize}
\item[](Gd$_{0.2}$La$_{0.2}$Nd$_{0.2}$Sm$_{0.2}$Y$_{0.2}$)CoO$_3$ \item[](Gd$_{0.2}$La$_{0.2}$Nd$_{0.2}$Sm$_{0.2}$Y$_{0.2}$)CrO$_3$
\item[](Gd$_{0.2}$La$_{0.2}$Nd$_{0.2}$Sm$_{0.2}$Y$_{0.2}$)FeO$_3$
\end{itemize}

Henceforth, the  mixed A-site (Gd$_{0.2}$La$_{0.2}$Nd$_{0.2}$Sm$_{0.2}$Y$_{0.2}$) will be denoted as (5A$_{0.2}$).

Powder X-ray diffraction (XRD) patterns were recorded using a Bruker D8 diffractometer with Bragg-Brentano geometry using Cu-K$\alpha$ radiation with a Ni filter. TOPAS 5 refinement software was used for Rietveld analysis of the XRD patterns. (A$_{0.2}$)CoO$_3$ and (A$_{0.2}$)FeO$_3$ crystallized into single-phase orthorhombic ($Pbnm$) perovskites \cite{Sarkar2017a}. 
In (A$_{0.2}$)CrO$_3$,  a small amount (2.3\,wt.\%) of secondary non-perovskite type (bixbyite, Gd$_{2}$O$_{3}$) phase could be observed.


Gd$_{2}$O$_{3}$ is paramagnetic in the range of temperatures investigated here \cite{Moon1975} so that any RE magnetism is carried by Nd and Sm.  We shall see that the magnetic state depends sensitively on which TM ion is considered, even seemingly that of the RE sublattice.

In a previous publication\cite{Sarkar2017a}, the structural properties of these compounds were systematically studied. The effect of the differing ionic sizes of A and B ions in perovskites are described by a tolerance factor $t$, where the ideal structure free of distortion is  $t = 1$. In the case of mixtures of ions on the A and B sites, an average radius is used (see \cite{Sarkar2017a} for details) to calculate $t$. The resulting distortions can be described by  the metric strain $\epsilon$, and two parameters expressing the distortion and tilting of the octahedra, expressed as a change in the coordination number ECoN(A) and ECoN(B) for A and B sites respectively. In\, \onlinecite{Sarkar2017a}, both the A(5B$_{0.2}$)O$_3$ and (5A$_{0.2}$)BO$_3$ type of high entropy variations were studied. It was found that both $\epsilon$ and ECoN(A) varied roughly linearly and ECoN(B) was constant with $t$.   

The one exception in this work, (5A$_{0.2}$)MnO$_3$ had a very different ECoN(B). In  (A$_{0.2}$)MnO$_3$, no secondary non-perovskite phase could be detected in XRD. However the XRD line shape of this sample  was highly asymmetric at room temperature \cite{Sarkar2017a}, and this asymmetry disappeared smoothly with increasing temperature.  The exact origin of this behaviour  is unclear; nevertheless one of the possible reason can be related to the Jahn-Teller effect of Mn$^{3+}$ cation \cite{Sarkar2017a}. Therefore,  this sample  will  be the subject to a separate study.

Structural details of all these systems are tabulated in the Supplementary Information Table S1. Included are space groups of primary and secondary phases as well as lattice parameters from diffraction data and Rietveld treatments. The included high resolution transmission electron microscopy (HR-TEM) data images confirm the XRD data. 

\subsection{Electronic characterization: XANES}\label{sec:exp_xanes}

It was possible to study the  valence states 
using X-ray absorption near edge spectroscopy  (XANES) at the UE46$\_$PGM-1 high field Diffractometer end-station on BESSY II of the Helmholtz-Zentrum Berlin \cite{Weschke2018}. XANES measurements were made at the M$_{4,5}$ edges of La, Sm, Nd and Gd  as well as the L$_{2,3}$ edges of Cr and Co, all at 15\,K. The signal was measured in total electron yield mode in an external magnetic field of 50\,mT. The sample powders were pressed into Indium foil to ensure electrical grounding, necessary for these highly insulating oxide powders. Because of the presence of many atomic species in these samples (including sample holder), careful subtraction of the background signal was necessary. More fundamentally, because of the computational difficulty of calculating the XANES edges from first principles in these mixed oxides, we have restricted our analysis here to comparisons with well established literature results. A recent review of the possibilities of XANES studies has been given by Henderson et al. \cite{Henderson2014} and in a larger context by Wende \cite{Wende2004c}. In addition, XMCD measurements for the (A$_{0.2}$)CoO$_3$ have been performed at the Co L$_{2,3}$ edges in a field of 6\,T, while the circular polarization was changed with the undulator.

The rare earth results will be presented below as one block, while the results for the transition metals  will be presented in the respective  subsections.

\subsection{Magnetic characterization and M\"ossbauer spectroscopy}\label{sec:exp_magn}

Magnetic characterization of the samples was performed using a Quantum Design MPMS3 Superconducting Quantum Interference Device (SQUID) vibrating sample magnetometer (VSM) with an additional furnace option. The sample mass was carefully measured and the samples were mounted in a dedicated Quantum Design powder sample holders with a brass sample holder stick. All magnetization measurements were done in VSM mode for the low temperature measurements. Temperature dependent measurements between 1.8  and 400\,K were performed following a zero-field cooled (ZFC) - field cooled (FC) protocol: The sample was cooled in zero magnetic field down to 1.8\,K. Then the external field $\upmu_0 H$ was applied and the magnetization then measured during warming up to 400\,K (ZFC branch). Subsequently, the magnetization was measured with the magnetic field applied from 400\,K to 2\,K (FC branch). Magnetic field dependent $M$($\upmu_0 H$) measurements at different temperatures were also performed after cooling in zero magnetic field.  High temperature magnetization measurements were performed for one sample using the Quantum Design oven. The powder was fixed on the heater stick with a MgO based cement and high $T$ measurements were performed in vacuum. In order to obtain ZFC data from above the magnetic transition temperature, samples were first heated up to 900\,K then cooled to 10\,K (lowest reachable temperature with the furnace) before warming in the field up to 900\,K and subsequent field-cooling to  10\,K.

$^{57}$Fe M\"ossbauer spectroscopy (MS) was carried out employing a $^{57}$Co:Rh source in transmission geometry using a triangular sweep of the velocity scale.  As it is conventionally done, all center shifts are given relative to $\alpha$-Fe at room temperature. High temperature measurements were performed with a Wissel  M\"ossbauer furnace.

\section{Results and discussion}\label{sec:results}

Multi-component samples with (Gd$_{0.2}$La$_{0.2}$Nd$_{0.2}$Sm$_{0.2}$Y$_{0.2}$)BO$_3$ with B = Co, Cr or Fe 
have been studied.  Subsections are organized with respect to the transition metal implemented in the compounds, which is in all cases decisive for the magnetic properties. 

First we present the results for the XANES spectra of the rare earths. We studied the four M$_{4,5}$ edges of La, Sm, Nd and Gd for  (5A$_{0.2}$)CoO$_3$ and   (5A$_{0.2}$)CrO$_3$.
These are shown in the supplementary materials, Fig. 4.
The RE 3d-4f absorption spectrum is characterized by two structures (denoted M5 and M4) which are separated in energy by a strong spin-orbit interaction. In all cases, the results did not differ between these  two compositions, indicating no change in the rare earth 3+ valance or spin-orbit interaction from one sample to the next. These results are in good agreement with literature results for the 3+ state. \cite{DeNadai2004, Kaya2018, Yasui2015, Tripathi2018}



\subsection{(5A$_{0.2}$)CoO$_{3}$}


The Co system was chosen because it was expected that the  cobalt would be in a Co$^{3+}$ low spin (LS) state with zero moment. Thus the remaining RE ions are probably the main contributors to the magnetic properties. This compound then serves as a reference to the other systems, where the TM ion, in addition, carries a finite moment. 
However, it is also known that in  $L$CoO$_3$ ($L$ = La, Pr, Nd),  Co$^{3+}$ can also transition to an intermediate (IS) or high spin (HS) state with decreasing $L^{3+}$ radius and increasing temperature\cite{Yan2004}.
Thus, it is important to determine experimentally the valance and spin state of Co.

\begin{figure}[htb]
   \[\includegraphics[width=1\columnwidth]{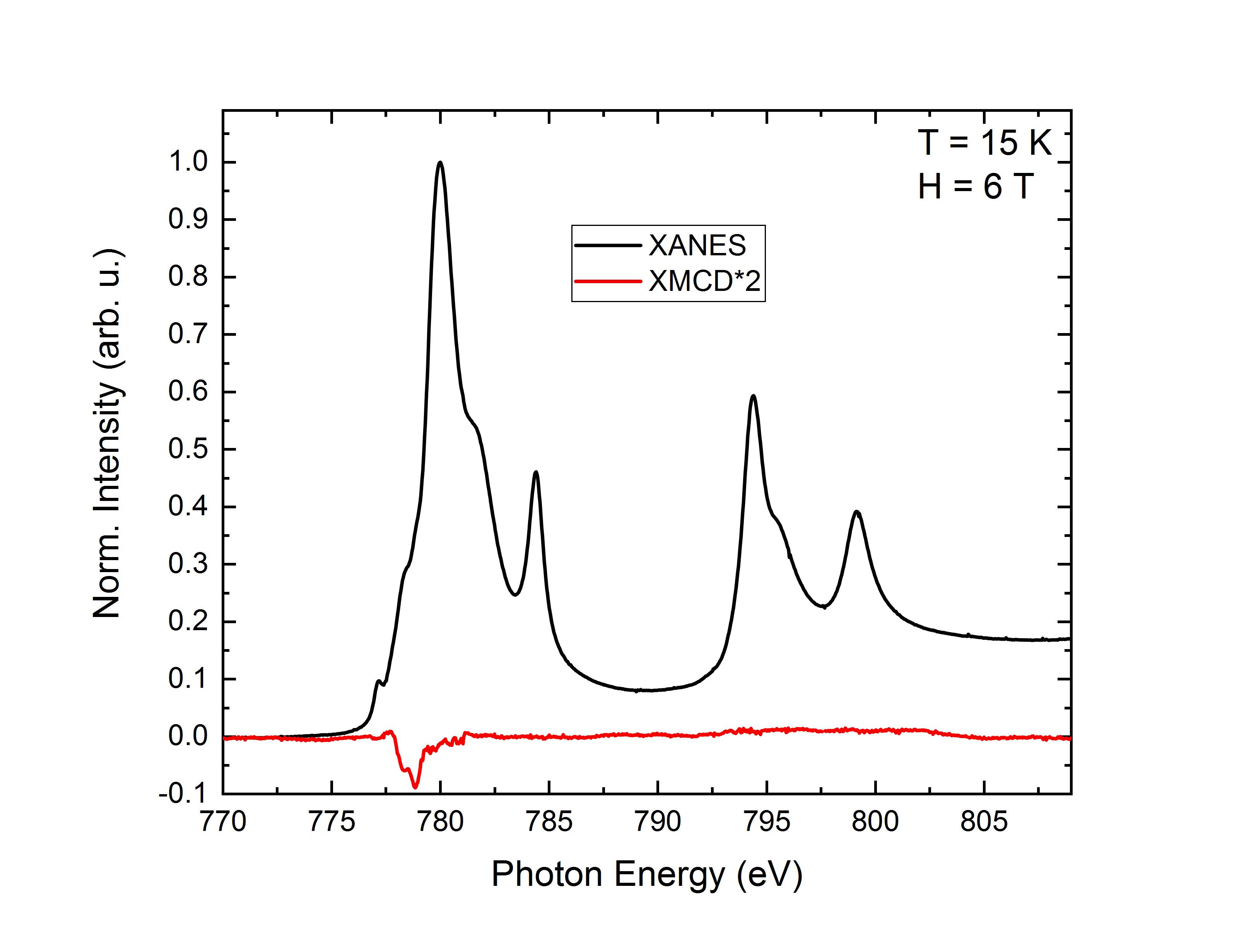}\]
\caption{XANES and XMCD L$_{2,3}$ spectra of Co  for the (5A$_{0.2}$)CoO$_3$  sample at 15\,K and in an external field of 6\,T. The spectra have been normalised to the edge jump. The XMCD spectrum is shown with a scale of 2. Near 785 and 800\,eV, we also see the M$_{4,5}$ XANES spectrum of Barium,  due to a very small Ba impurity (estimated from the signal to be much less than 1 at. \%).
}    
\label{fig_XANES_Co}
\end{figure}

\begin{figure}[htb]
   \[\includegraphics[width=1.1\columnwidth]{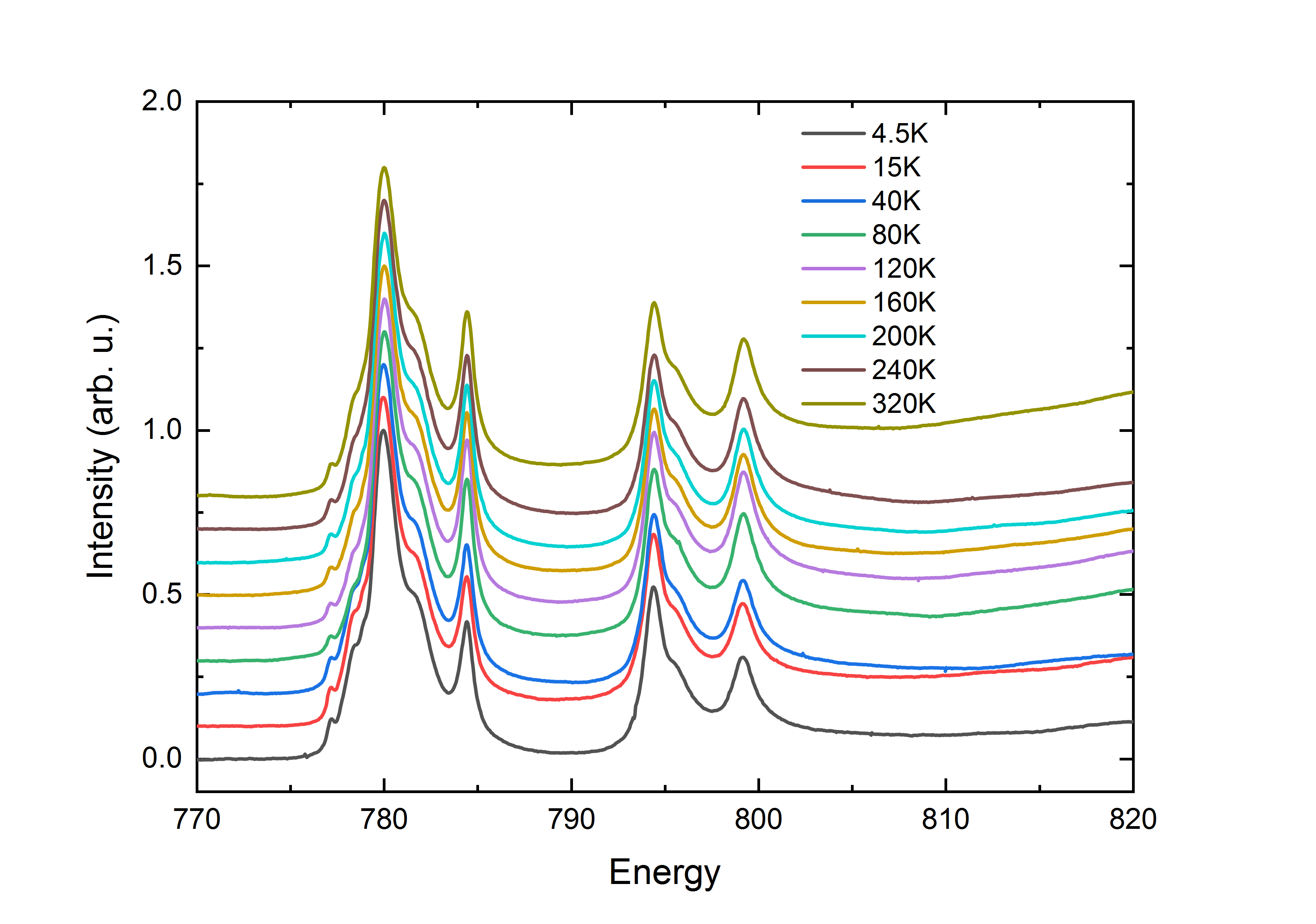}\]
\caption{XANES  L$_{2,3}$ spectra of Co  for the (5A$_{0.2}$)CoO$_3$  sample for temperatures between 4.5 and 320\,K. The spectra have been normalised to the maximum of the Co L$_{2,3}$ whiteline intensity.  Near 785 and 800\,eV, we see the M$_{4,5}$ XANES spectrum of Ba, found in all of our cobalt containing samples, and is due to a small Ba impurity.  The Co L$_{2,3}$ edges do not change in this temperature range showing the suppression of the LS-HS transition.
}    
\label{fig_XANES_Co_T}
\end{figure}

Fig. \ref{fig_XANES_Co} shows both the XANES and XMCD spectra of the (5A$_{0.2}$)CoO$_{3}$ in the energy region of Co L$_{2,3}$. In addition to the cobalt L$_{2,3}$ lines, two lines are seen which have been identified as the L$_{2,3}$ lines of barium, an minute impurity (less than 1 at. \%) in the cobalt precursor. The Co L$_{2,3}$ peaks show clear structure of Co$^{3+}$ low spin (LS) but with a small induced moment (seen in the XMCD signal) parallel to the external magnetic field. The XANES scans made at temperatures between 4.5 and 320\,K, Fig. \ref{fig_XANES_Co_T}, show no sign of any valance change, so that we can conclude that the cobalt remains  Co$^{3+}$ low spin (LS), with a small induced moment due to the magnetic environment (the surrounding RE ions). Unfortunately in a strongly non-collinear system, it is difficult to determine the magnitude of the magnetic moments.

\begin{figure}[htb]
\[\includegraphics[width=1\columnwidth]{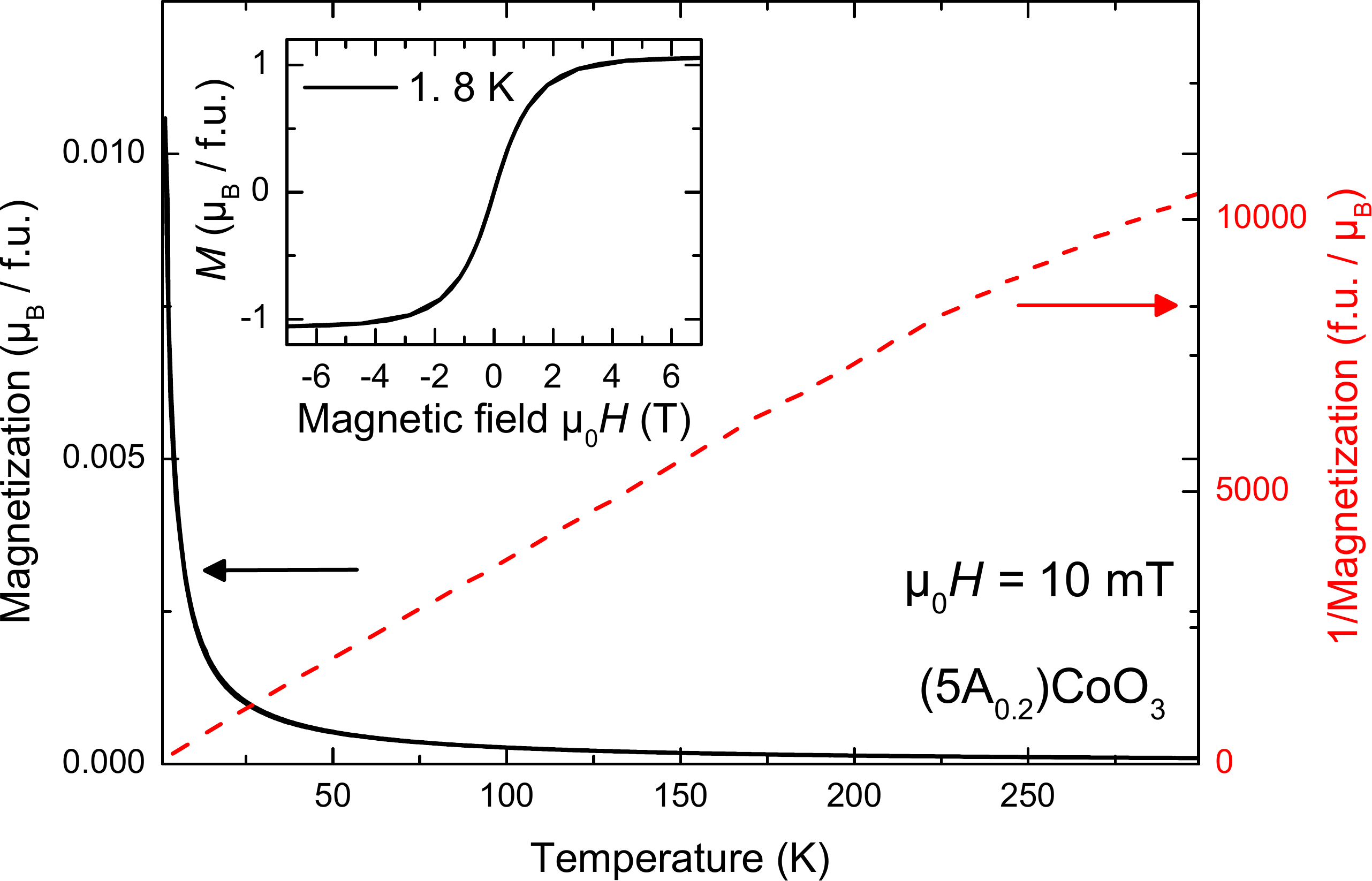}\]
\caption{Magnetization and inverse magnetization of (5A$_{0.2}$)CoO$_{3}$ as function of temperature. The inset shows magnetic field dependent measurement at 1.8\,K.}
\label{fig_m_aco}
\end{figure}

The magnetization results for the (5A$_{0.2}$)CoO$_{3}$ compound are presented in Fig.\,\ref{fig_m_aco}.  Shown is the magnetization and its inverse (magnetic susceptibility) measured in a temperature window from 1.8 to 295\,K in a small external field of $\upmu_0H$ = 10\, mT.  The system  clearly shows  Curie-like paramagnetic behavior  down to 1.8\,K, as seen in  the inverse susceptibility (dotted red line), which is linear down to low temperatures, with no temperature offset.  This is in agreement with the magnetic behavior of other RE-Cobaltites, such as Gd, Sm, \cite{Ivanova2007}, La and Nd\cite{Yan2004}.  The insert shows the magnetization in fields up to 7 Tesla at 1.8\,K, which saturates at about 1 $\mu_B$ per formula unit.

\subsection{(5A$_{0.2}$)CrO$_{3}$}

\begin{figure}[htb]
\[\includegraphics[width=1\columnwidth]{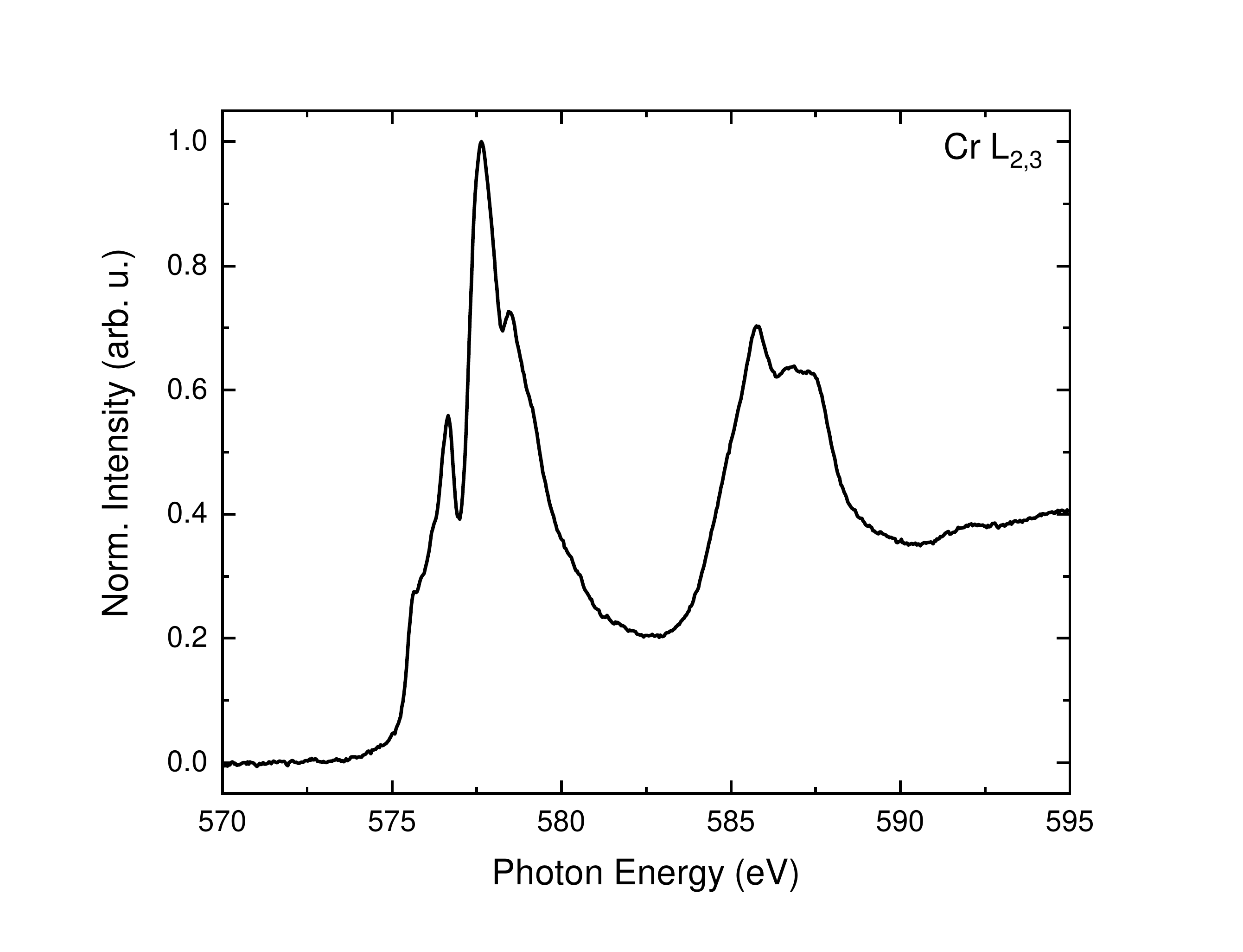}\]
\caption{XANES spectra of Cr at the  L$_{2,3}$ edges for the (5A$_{0.2}$)CrO$_3$   sample. The contribution at 592\,eV can be attributed to a second harmonic peak of the Gd M$_5$ edge.}
\label{fig_XANES_Cr}
\end{figure}

The XANES spectrum at the L$_{2,3}$ edge of Cr in (5A$_{0.2}$)CrO$_{3}$ is shown in Fig. \ref{fig_XANES_Cr}. Seong et al. \cite{Seong2018} have published XMCD/XAS spectra for a wide range of Cr-containing oxides. They compare half-metallic CrO$_2$ to trivalent Cr$_2$O$_3$ and Cr-doped Al$_2$O$_3$\  as well as Cr metal.  Their results for half-metallic CrO$_2$ have been explained by a mixture of Cr$^{3+}$ and Cr$^{4+}$ (their figure 3 b), and very different from the other Cr valance state results presented in their publication. They have explained the magnetic properties of CrO$_2$ as resulting from double exchange (DE) between Cr$^{3+}$ and Cr$^{4+}$ ions.

Our results actually show somewhat less broadening than theirs, but basically we can conclude on a similar mixed valent state as found in half-metallic ferromagnetic CrO$_2$. We thus conclude that at least the Cr - Cr interactions in this mixed oxide are characterized by DE. How Cr$^{4+}$ ions are stabilized in this HEO structure is an open question. One possibility is cation vacancies, as charge compensation vie the RE is not possible (and as we have seen, they are all 3+). 

\begin{figure}[htb]
\[\includegraphics[width=1\columnwidth]{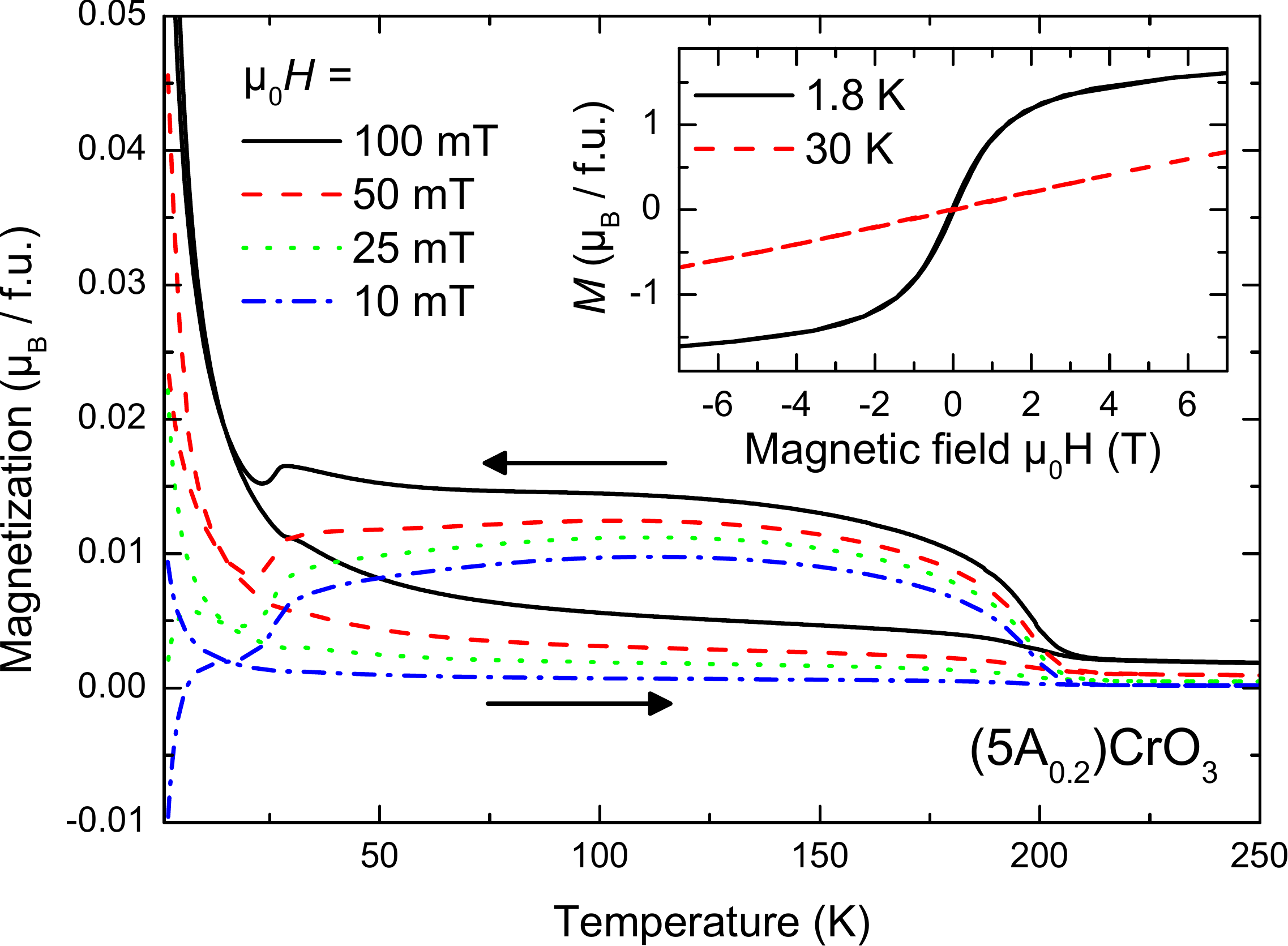}\]
\caption{Magnetization of (5A$_{0.2}$)CrO$_3$ as function of temperature, measured after zero field cooling and field cooling in different magnetic fields. The inset shows magnetic field dependent measurement at 1.8 and 30\,K .}
\label{fig_m_acr}
\end{figure}

The magnetic state of the (5A$_{0.2}$)CrO$_{3}$ compound shows a complex behavior for the ZFC and FC branches, very different  from that of  (5A$_{0.2}$)CoO$_{3}$. Temperature dependent $M(T)$ curves shown in Fig.\,\ref{fig_m_acr} indicate a  magnetic transition at about 198\,K.
But there is a large hysteresis between the ZFC and FC branches. In both branches, there is in addition dramatic shifts in M(T). First at very low external fields (10 mT) in the ZFC branch, the moment starts out small and almost disappears on heating. The low field FC branch  shows initially a much larger moment, but surprisingly reverses in direction at low temperature in two steps. At higher external fields (100 mT), both ZFC and FC branches lead to large magnetization at low temperatures, but this increase disappears at around 25 K.

The most probable magnetic structure is that of a canted antiferromagnet (AFM). This proposed structure would lead to a small net moment below the transition, as well as a Curie-Weiss form for the inverse susceptibility above (negative intersection with temperature axis at about -100\,K). Canted AFM structures have been also found for most of the binary rare RE-chromites \cite{Shamir1981,Sharma2014,Prado-Gonjal2013,Sardar2011}. 

The low temperature moment appears to be the effect of the RE system freezing magnetically. 
At very low external fields, this results in a negative moment.
That the FC branch reverses magnetization direction shows that in the range from the magnetic transition temperature down to ca. 30 K, the magnetic state is dominated by the Cr ions. At lower temperatures, the RE system orders antiparallel to the Cr system. Since in the FC branch at higher fields, the Cr system has been polarized in the direction of the external field, the RE system polarizes in the negative direction. This unusual result is maintained by the crystalline anisotropy and antiferromagnetic RE-Cr interactions, keeping the Cr system polarized in the positive, and the RE system in the negative direction. At higher external fields, the crystalline anisotropy is overcome and the polarization is always positive, and dominated at low temperatures by the RE part.

Thus we observe  a spin-reorientation transition below 30\,K. 
Spin reorientation occurs  in other RE-orthochromites (Sm, Gd, Nd) as well \cite{Hornreich1978, Rajeswaran2012a}. In these cases, the canted AFM structure of the Cr$^{3+}$ spin lattice rotates into the plane, resulting in zero net magnetization. However in most of the cases reported in the literature, magnetization reversal is not observed, as is here the case.



This is a reasonable assumption as some of the binary compounds show AFM coupling between the rare earth system and the Cr spins \cite{Shamir1981}. For example, a magnetization reversal  has been observed in GdCrO$_{3}$ and is attributed to the antiparallel coupling of Gd$^{3+}$ and Cr$^{3+}$ ions\cite{Cooke1974a}, while the other magnetic RE compounds with Sm and Nd do not show this behavior \cite{Wang2015}. It is thus a reasonable assumption  that also in the case of the mixed RE sites it is mostly the coupling of the Gd$^{3+}$ moment to the Cr spin lattice, which leads to the observed low temperature magnetization reversal behavior.
 Field dependent measurements at 1.8\,K show the presence of the magnetic moments of the RE ions. However due to the weak coupling of the rare earth moments to the Cr sublattice, or the presence of  highly frustrated magnetic exchange interactions, saturation is not reached.

\begin{figure}[htb]
\[\includegraphics[width=1\columnwidth]{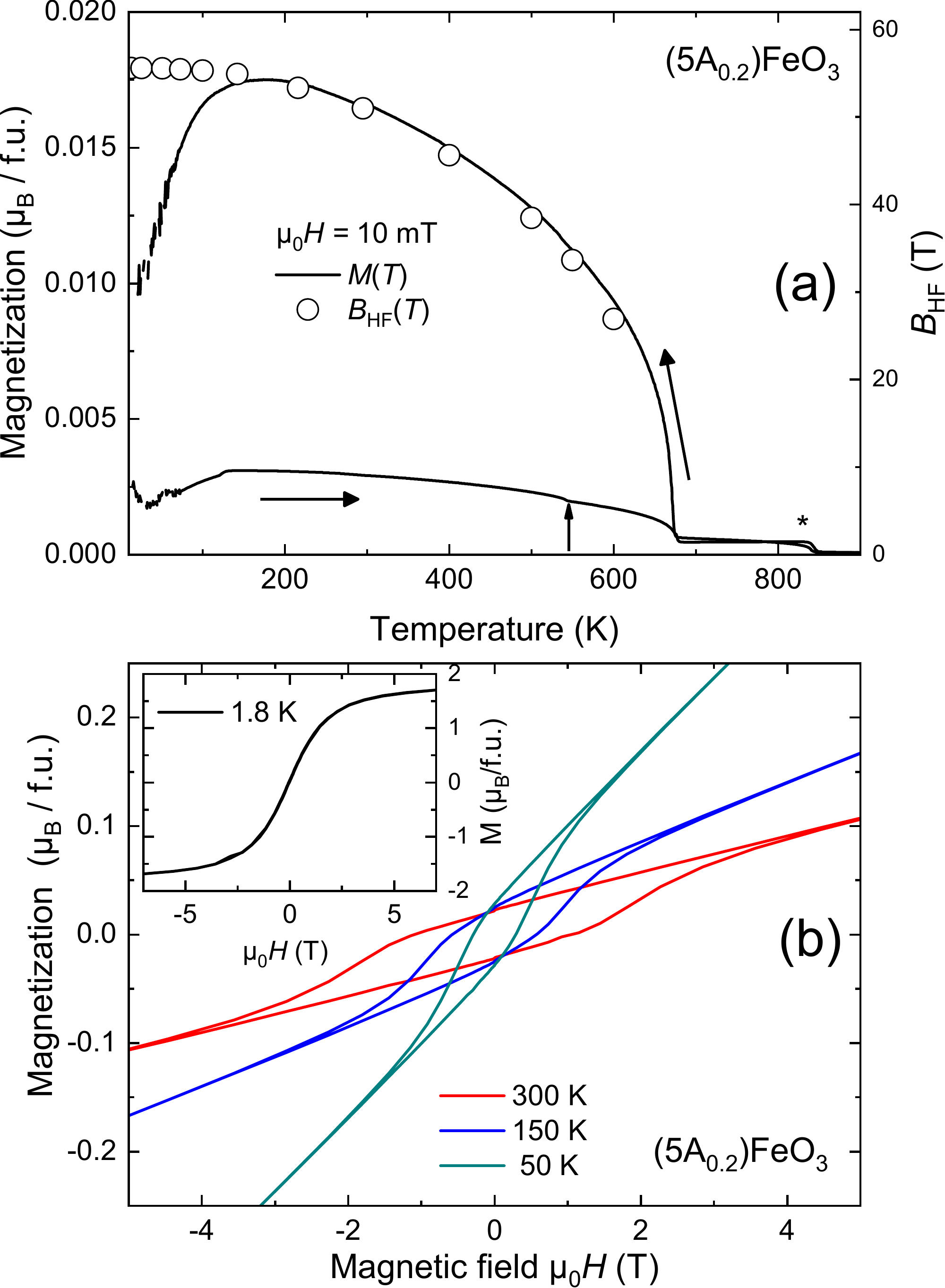}\]
\caption{(a) Magnetization as function of temperature. ZFC (after cooling in zero field from 900\,K) and FC magnetization of (5A$_{0.2}$)FeO$_3$ as function of temperature from 15 to 900\,K (solid line). The small arrow points at a minute magnetic transition, the asterixs indicates the position of a high temperature magnetic transition of a Fe-oxide impurity. Magnetic hyperfine field $B_{\mathrm{HF}}$ from M\"ossbauer spectroscopy (open circles). (b) Magnetic field dependent measurement at 300, 150 and 50\,K. The inset shows the same measurement at 1.8\,K }
\label{fig_m_afe}
\end{figure}

\subsection{(5A$_{0.2}$)FeO$_{3}$}

From the M\"ossbauer studies (presented below), we know that Fe is in a 3+ HS state. Thus we did not study the iron edge in the XANES studies since they would have only have confirmed this.

First we present the SQUID magnetization studies.
The (5A$_{0.2}$)FeO$_{3}$ compound shows the highest magnetic transition temperature of about 675\,K observed in the series of samples. Results of magnetometric characterization are summarized in Fig. \ref{fig_m_afe}. The observed small magnetization points again to an AFM arrangement of the Fe sublattice, as expected from the binary oxides \cite{Eibschutz1967a}, while the observed magnetization stems from the slight canting of the Fe$^{3+}$ spins as observed in nearly all orthoferrites, which possess a $G$-type canted magnetic structure \cite{Bousquet2016}.

The ZFC curve is measured after cooling in zero-field from above the $T_{\mathrm{N}}$. This branch is comparable in magnitude to that found for Cr. 
The large difference between ZFC and FC branch shows the difficulty in aligning the canted magnetic moments with the small magnetic field applied, which points towards the presence of a large magnetocrystalline anisotropy. However cooling in a magnetic field from above the  $T_{\mathrm{N}}$ leads to a net magnetization which is much larger due to the alignment of the canted moments (FC branch). Yet below about 100\,K a decrease of the magnetization is observed, which can be related to a spin-reorientation transition of the Fe magnetic sublattice, similar to the one observed in NdFeO$_{3}$ below 130\,K\cite{Sawinski2005} but possibly also to an antiferromagnetic ordering of some of the RE magnetic moments with respect to the canted FM of the Fe sublattice as was the case for the Cr sample.  

An increase of the magnetization at low temperatures due to slow relaxation of the large rare-earth magnetic moments could not be observed as the measurements  were limited to  15\,K and above due to the  furnace setup used. However, measuring in a regular setup to lower temperature shows the PM-like signal of the RE magnetic moments below 10\,K(not shown), as also seen in the Langevin like appearance of the  $M(\upmu_0H)$ measurement at 1.8\,K in the inset of Fig.\,\ref{fig_m_afe}(b). A small change in slope of the magnetization curve (ZFC branch)  is observed at 550\,K of unknown origin (arrow in the figure). The second magnetic transition observed at even higher temperatures of 840\,K is most likely related to small impurities in the sample of Fe-bearing oxides, which have high transition temperatures such as e.g. magnetite. 

\begin{figure}[htb]
\[\includegraphics[width=1\columnwidth]{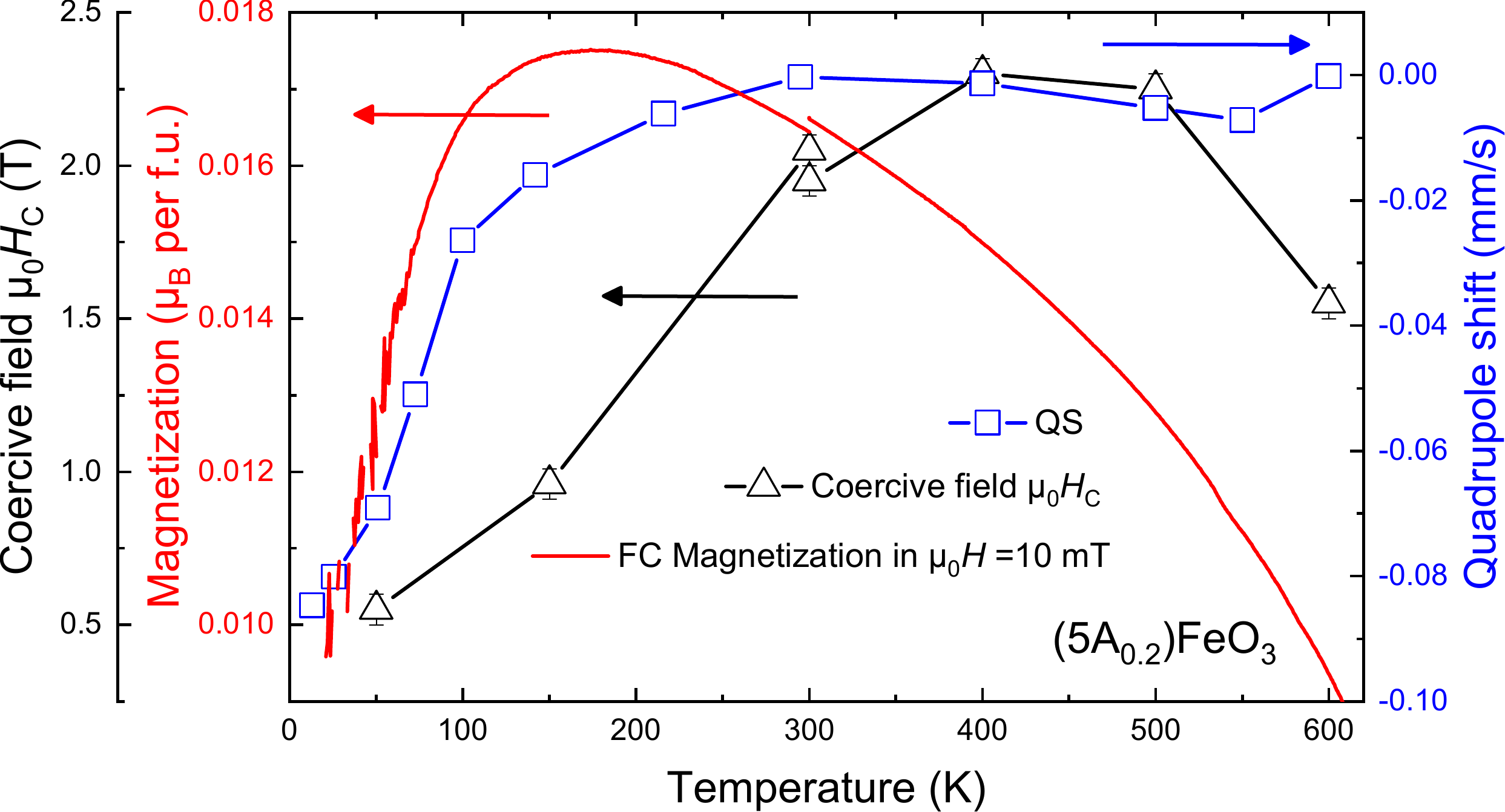}\]
\caption{Synopsis of temperature dependence of the coercive field $\mu H_{\mathrm C}$, magnetization in field.cooling and the quadrupole coupling constant of the magnetic sextet determined from  $^{57}$Fe M\"ossbauer spectra recorded at temperatures from 13 to 700\,K.}
\label{fig_Hc_QS}
\end{figure}

Fig.\,\ref{fig_m_afe}(b) also features $M(\upmu_0H)$ measurements at 50, 150, and 300\,K.  From these and further hysteresis measurements the coercive field $\mu H_{\mathrm{C}}$ of  the FM component was determined after subtracting the linear part of the magnetization curve and the results are compiled in Fig.\ref{fig_Hc_QS}. Surprisingly, a distinct increase of the coercive field  is observed with increasing temperature, reaching up to about 2\,T at 300\,K. This is a quite anomalous temperature dependence, as usually magnetocrystalline anisotropy decreases with temperature \cite{Zener1954}. Therefore, it may be related to a spin reorientation transition of the canted moments, as observed in TmFeO$_3$ \cite{Wolfe1967} or in SmFeO$_3$ \cite{Babu2016,Lee2011a}. However, the observed magnitude of the coercive field $\upmu_0H_{\mathrm{C}}$ of about 2\,T in the present HEO ferrite samples, is huge compared to what has been observed in the above cited literature work, which reached a few mT only. A comparable high coercive field was only observed in a highly distorted LuFeO$_3$ compound \cite{Zhu2012a}. In the following results from element specific M\"ossbauer spectroscopy will be used to shed some light on the obtained magnetization data.

\begin{figure}[htb]
\[\includegraphics[width=1\columnwidth]{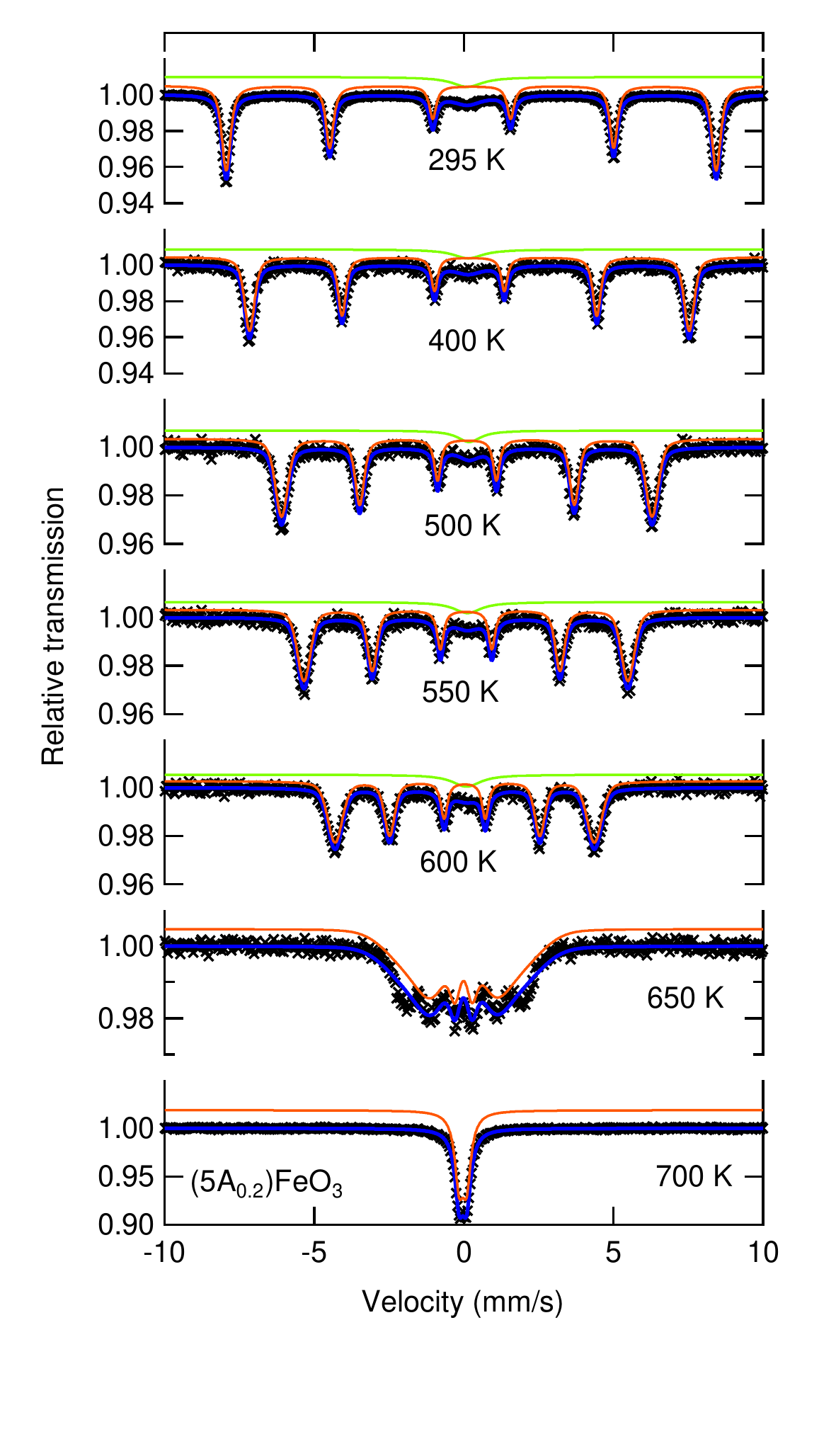}\]
\caption{$^{57}$Fe M\"ossbauer spectra recorded at various temperatures across to the Neel temperature of the Fe sublattice. The singlet originates from the M\"ossbauer furnace.}
\label{fig_ms_afe}
\end{figure}

M\"ossbauer spectroscopy investigations performed at temperatures from 13\,K to above the transition show a decrease of the magnetic hyperfine field $B_{\mathrm{HF}}(T)$ with increasing temperature (see Fig.\,\ref{fig_m_afe}) in accordance with the magnetization data. 
Additional low temperature spectra are presented in the supplementary material.
As can be seem from the low temperature behavior of $B_{\mathrm{HF}}$, we do not see a decrease in $B_{\mathrm{HF}}(T)$ at low temperature, which excludes a reduced local Fe magnetic moment as possible explanation of the magnetization data. However, the quadrupolar electric field gradient (EFG) line shift 2$\epsilon$ in the magnetically split spectra (see Fig.\,\ref{fig_Hc_QS}, right ordinate) changes systematically with temperature and may be well correlated with the proposed spin reorientation transition. 

The shift in line position described by the electric field gradient interaction depends on the angle between the principle axis of the EFG  $V_{zz}$ and the direction of the nuclear hyperfine field $B_{\mathrm{HF}}$ together with the magnitude of the main component of the EFG \cite{Gutlich2011}. 
The line shift is given by:
\begin{equation}
    \begin{split}
        \epsilon(m) &= (-1)^{|m|+1/2} (eQV_{zz}/8) (3 cos^2(\Theta)-1 \\ 
                    &+ \eta sin^2(\Theta)cos(2\phi))
    \end{split}
\end{equation}
where $m$ is the magnetic quantum number, $Q$ the nuclear quadrupole moment and $\eta$ the asymmetry of the EFG tensor, all of the  of the 3/2 nuclear state. The angles $\Theta$ and $\phi$ express the direction of $B_{\mathrm{HF}}$ in the coordinate system of the EFG tensor (see for example \cite{Gutlich2016}).


Because of the angular dependence, a spin reorientation will lead to changes in the EFG line shift $2\epsilon$. The direction of the observed shift is towards larger (negative) values at lower temperature, where the RE system polarizes. The RE-Fe antiferromagnetic exchange interactions then affect the directions of the Fe moments, decreasing the  magnetization.
A similar feature in the quadrupole coupling as a function of temperature has been observed in the spin reorientation transitions of CeFeO$_3$\cite{Robbins1969}
and SmFeO$_3$\cite{Eibschutz1967} orthoferrites, which supports our conclusion that the two anomalies in magnetization measurements, low temperature decrease as well as increase of coercive field with temperature are correlated with electronic structure changes as seen by M\"ossbauer spectroscopy.

 Above the transition a doublet is observed, with no additional sextet. Hence, the small high temperature magnetic contribution seen in the magnetization measurement  stems from a minor magnetic impurity which is not visible in spectroscopy. The magnetic hyperfine field of 55.6(1)\,T at 13\,K,  is comparable to that for Fe$^{3+}$  found in other binary orthoferrites, which show a dependence of the $B_{\mathrm{HF}}$ on the ionic radii of the rare earth ions\cite{Eibschutz1967}. The present multicomponent sample  integrates into that series of parent compounds.  Moreover the spectrum is surprisingly narrow with only broadening despite the chemical disorder and possible structural distortion around the Fe site. It can be well represented by a Gaussian distribution of hyperfine fields with a standard deviation of 0.4\,T, This result is in agreement with our previous work on a La(Cr$_{0.2}$Co$_{0.2}$Fe$_{0.2}$Mn$_{0.2}$Ni$_{0.2}$)O$_3$ HEO perovskites \cite{Witte2017}, where the local Fe environment was also comparably well defined.

\subsection{Discussion}

In this section we compare the results of the present HEO perovskites with their parent compounds. In the parent orthochromite, -ferrite and manganite compounds it is well known that the rare earth ions act as a structural stabilizing agent, defining the degree of distortion of the lattice as described by the B-O-B bond angle and the octahedral tilting \cite{Goldschmidt1926}. This structural distortion is directly correlated to the magnetic properties, such as the  transition temperature $T_{\mathrm N}$\cite{Zhou2006,Zhou2010a}.
  
As given by the XANES results for the rare earths, Fig. 4 of the Supplementary Information, these ions retain their usual 3+ character in all of our samples.   For the transition metal ions, Co and Cr were measured, and we surprisingly found that Cr is in a Cr$^{3+}$ - Cr$^{4+}$  mixed valent state, also present in half-metallic CrO$_2$ but not in the other oxides of Cr. In this latter phase, Cr-Cr  pairs interact by double exchange. But we do expect that the RE ions will play the same role
in the magnetic properties
as they do in the parent compounds, including RE-TM antiferromagnetic superexchange.

\begin{figure}[htb]
\includegraphics[width=1\columnwidth]{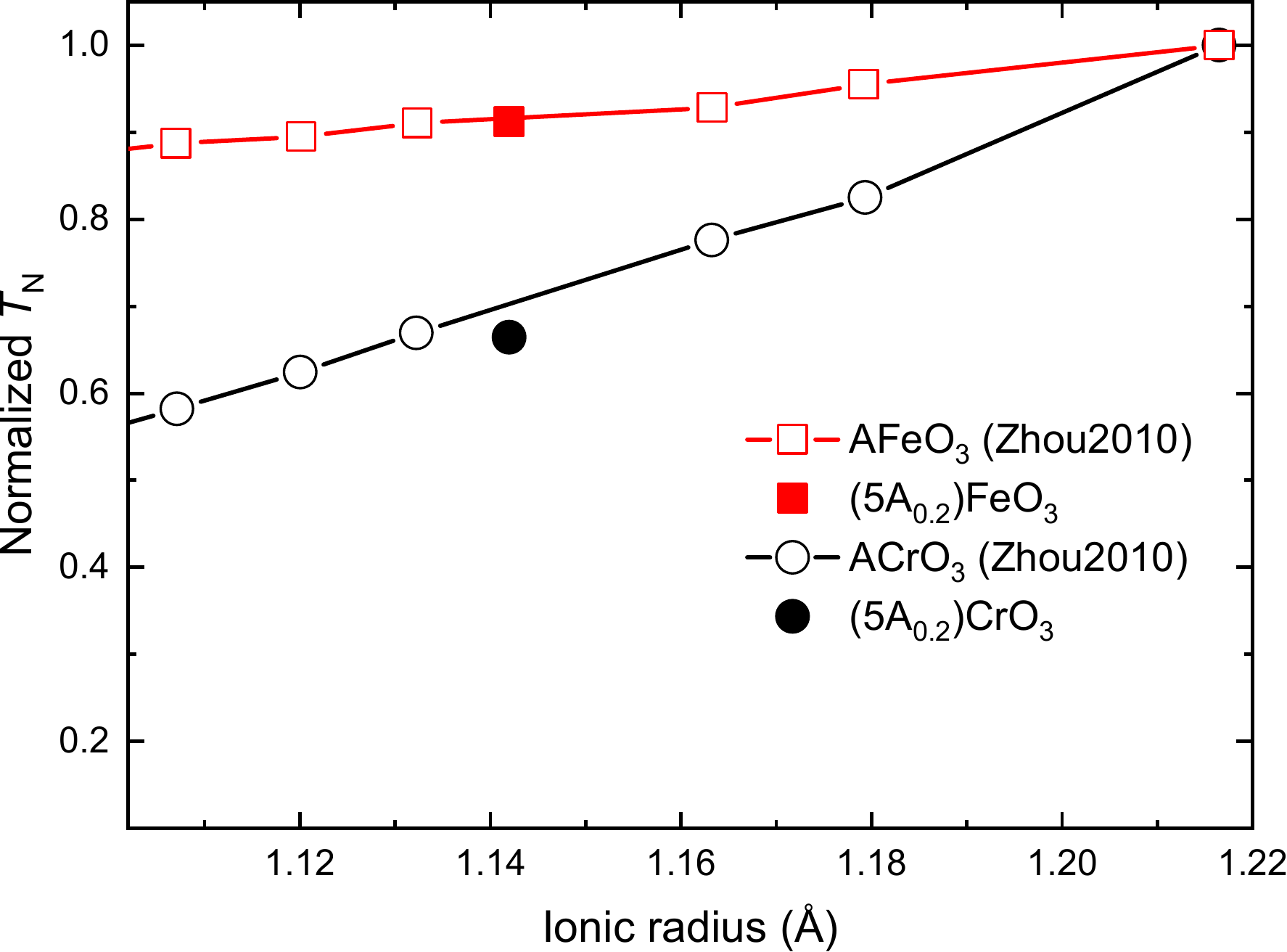}
\caption{Representation of the normalized $T_{\mathrm N}$ for the orthoferrites and orthochromites\cite{Zhou2010a} 
(open symbols) in comparison to the  $T_{\mathrm N}$  of the respective PE-HEO (full symbols).}
\label{fig_TNvsIR}
\end{figure}
  
In Fig.\,\ref{fig_TNvsIR} we compare the  $T_{\mathrm N}$ of the HEO (closed symbols) to those of the binary oxides in a structural series as a function of ionic radii (open symbols). The ferrite compound perfectly integrates into that series, which shows that the disordered nature of the A-site, and with that any local distortion of the lattice, has no significant impact on the strength  of the orbital overlap of the $d$-states of the Fe$^{3+}$ ions with the $p$ states of the O$^{2-}$ ions. That can be understood considering that the 3$d^5$ electronic configuration of Fe$^{3+}$ forms mostly a $\sigma$ bond with oxygen, which is less affected by a locally distorted B-O-B bond angle than a $\Pi$ bond\cite{Zhou2010a}.

The same argument can explain that in the case of the HEO orthochromite we observe a larger deviation from the overall linear behavior. In the Cr$^{3+}$ case also $\Pi$ bonding states contribute significantly to the orbital overlap which makes it more sensible to any distortion of the Cr-O-Cr bonds. This is effectively the reason for the stronger decrease of the $T_{\mathrm N}$ with decreasing ionic radius compared to the orthoferrites. It is thus a reasonable assumption that local distortions of the Cr-O-Cr bonds also affect the overall magnetic coupling of the system resulting in a relatively reduced transition temperature.


A further argument for the importance of local structural distortion in this system comes from our previous work on the structural aspects of these compounds \cite{Sarkar2017a}. There it was found that the the x-ray diffraction pattern needs to be represented with several perovskite phases with slightly different unit cell parameters. This is then in agreement with the present results showing a very complex magnetic state.

Finally, the (5A$_{0.2}$)CoO$_3$ compound shows a paramagnetic state like many of the parent binary oxides. However some of the cobaltites also show low temperature magnetic order of the rare earth magnetic moments\cite{Knizek2014}, which is not observed here possibly due to the chemical disorder on the A-sites.

\section{Summary and conclusion}\label{sec:sum}

In this article we provide comprehensive magnetic and spectroscopy characterization of new compounds of rare-earth orthocobalties, -chromites and ferrites,
which feature a mixture of five different rare-earth ions on the A-site, namely Y, La, Gd, Nd and Sm. These  compounds are members of the family of high-entropy oxide perovskites \cite{Sarkar2017a}, which have been recently introduced.

In general it was found that the magnetic properties of the  HEO perovskites   are strongly linked to the behavior of the binary oxide series. All of them, except for the cobaltite, are canted antiferromagnets with transition temperatures $T_{\mathrm N}$ that are linked to the structural distortion of the perovskite structure, as in the parent compounds. Yet, the agreement with the behavior of the binary oxides is best for the ferrites, and we observe stronger deviations for chromites.  
These differences are explained by the different orbital overlap in the B-O-B bond in 
chromite, which results in a more expressed sensitivity of their magnetic exchange coupling to a local structural distortion created by the mixed A-site. The cobaltite sample behaves paramagnetic down to 2\,K.
However, the individual compounds show interesting details in their magnetic characteristics, which are summarized in the following paragraphs.

The (5A$_{0.2}$)CrO$_3$ compound shows interesting features in the temperature dependent magnetization curves, which result in a magnetic field dependent magnetization reversal at low temperatures, accompanied by a possible spin-reorientation transition of the Cr sublattice. The magnetization reversible maybe well related to the coupling of the rare earth moment, which set in at low temperatures.

The (5A$_{0.2}$)FeO$_3$ has the highest $T_{\mathrm N}$ of the investigated materials. Interestingly it also shows signs of a spin-reorientation transition in the temperature dependent magnetization, but more significant is the large increase of the coercive field of the canted ferromagnetic moment  with increasing temperatures reaching a value of about 2\,t at ambient temperature. The spin-reorientation interpretation is supported by M\"ossbauer spectroscopy which shows a distinct change of the quadrupole coupling constant in the same temperature range, which can be correlated with the spin reorientation.


We summarize the points pertaining to high entropy oxides:
\begin{enumerate}
\item The crystalline stability of the mixed rare-earth transition metal perovskites in the high entropy regime makes them interesting candidates for possible applications. 
\item We have shown that the valance instability in cobalt containing perovskites is suppressed.
\item The upper AF transition temperature depends on the average ionic radius similar to the case in the pure substances.
\item The magnetic behaviour of the (5A$_{0.2}$)BO$_3$ series depends sensitively on the nature of the TM on the B site. 
\item At lower temperatures, the magnetic states in the TM and RE subsystems interact strongly.  
\item For B = Fe, M\"ossbauer spectroscopy revealed the effect of the magnetic ordering on the RE sublattice, moreover a large coercive field of 2\,T at room temperature was observed.
\end{enumerate}

As such the family of rare-earth transition metal perovskites display a vast research space for further studies, including new members such as e.g. orthovanadates, -scandates or cuprates to name a few. Moreover, a delicate choice of the rare-earth elements may allow for the design of novel properties. To this respect also the use of mixed valence rare-earth elements is highly promising and may induce unknown physical effects
like it was observed in the rare-earth based fluorite HEO\cite{Sarkar2017b}.

\begin{acknowledgments}
  We acknowledge  financial support from the Helmholtz Association and the Deutsche Forschungsgemeinschaft (DFG), project HA 1344/43-1 and WE 2623/14-1. 
  The authors thank the Helmholtz-Zentrum Berlin for access to synchrotron radiation facilities  for the experiment at the high field end-station  UE46\_PGM-1 beamline, project 191-07848-ST.
\end{acknowledgments}

%
%
%
%

%

\end{document}